\documentclass[a4paper,openbib,10pt]{article}
\usepackage{graphicx}
\usepackage{float}
\usepackage{layout}

\begin{document}
\renewcommand{\thefootnote}{\fnsymbol{footnote}}
\setlength{\bibindent}{0.25in}
\renewcommand{\figurename}{\small \textbf{Figure}}

\vspace*{96pt}

{\Large\bf \flushright
NONEQUILIBRIUM QUASIPARTICLES \\
AND ELECTRON COOLING BY \\
NORMAL METAL -- SUPERCONDUCTOR \\
TUNNEL JUNCTIONS\\ }
\vspace{24pt}
{\large \flushleft Dmitri Golubev and Andrei Vasenko
\footnote{ Dmitri Golubev, Institut f\"ur Theoretische Festk\"orperphysik, Universit\"at Karlsruhe,
\protect\hbox{D-76128} Karlsruhe, Germany. Andrei Vasenko, Department of Physics, Moscow State University, Vorobjovy Gory,
119992 Moscow, Russia.}}
\vspace{24pt}


\thispagestyle{plain}

\section{Introduction}

\pagestyle{myheadings}
\markboth{\rm D. GOLUBEV AND A. VASENKO}{\bf NONEQUILIBRIUM QUASIPARTICLES \dots}

It is known that Normal metal -- Insulator -- Superconductor (NIS) tunnel junction
in a certain range of bias voltages
cools the normal metal electrode. Within
a simple ``semiconductor model'' of a superconductor one can derive
a cooling power of a single NIS junction \cite{Pekola1}:
\begin{equation}
P=\frac{1}{e^2R}\int dE\, \frac{\theta(E^2-\Delta^2)|E|}{\sqrt{E^2-\Delta^2}}\,(E-eV)\,[f_N(E-eV)-f_S(E)].
\label{P0}
\end{equation}
Here $e$ is a positive absolute value of the electron charge, $R$ is the normal state
resistance of the NIS junction, $V$ is the bias voltage, $E$ is the energy of
quasiparticles in the superconductor, $\Delta$ is the superconducting gap.
$f_N$ and $f_S$ are distribution functions in normal metal and superconductor
respectively. In equilibrium these are Fermi functions. The cooling power
(\ref{P0}) turns out to be positive if $V<\Delta/e.$

Microrefrigerator, based on a NIS tunnel junction, has been first
fabricated by Nahum and Martinis \cite{Nahum}. They have used a
single NIS tunnel junction in order to cool a small normal metal
strip. Later Leivo {\it et al} \cite{Pekola1} have noticed that
the cooling power of a NIS junction (\ref{P0}) is an even function
of an applied voltage, and have fabricated a refrigerator with two
NIS junctions in series. By doing so they have achieved much
better performance of the microrefrigerator. However, they have
also observed a sharp drop of the cooling power at the base
temperature below 200 mK. This drop has been attributed to the
heating of the superconducting electrode, which absorbs both the
cooling power (\ref{P0}) and electric power $VI$ \cite{Jochum}.
Pekola {\it et al} \cite{Pekola2} have demonstrated that this
problem can be solved if one covers the superconducting electrode
by an additional layer of normal metal, which serves as
quasiparticle trap and removes excited quasiparticles from the
superconductor. The contact between the superconductor and the
trap can be either direct or through an oxide layer.

\begin{figure}
\begin{center}
\includegraphics[width=11cm]{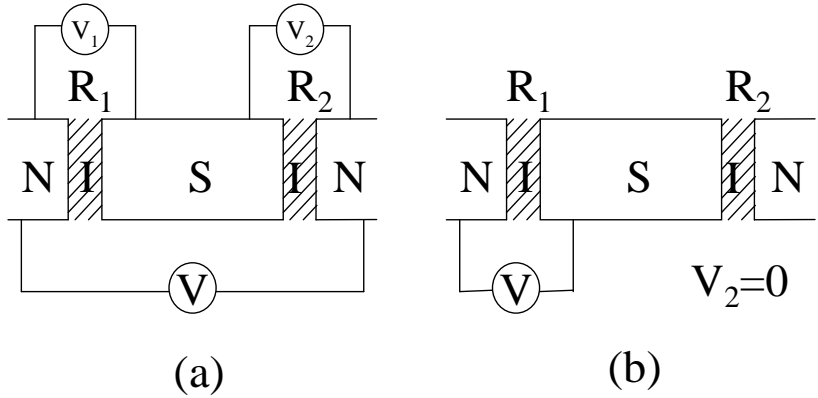}
\end{center}
\caption{System of two NIS junctions with normal state resistances
$R_1$ and $R_2,$ $R_1>R_2.$ The junction number 1 cools the left
normal electrode. The right normal electrode is a quasiparticle
trap. Two possible bias methods are shown: (a) the bias voltage is
applied between the normal metal leads, $V_1\not=0,$ $V_2\not=0$
and $V=V_1+V_2;$ (b) the bias voltage is applied between the left
normal lead and the superconductor, while the voltage drop at the
second junction is zero, in this case $V=V_1$ and $V_2=0.$ }
\end{figure}
The aim of this contribution is to consider processes in the
superconducting electrode in more detail and obtain some
quantitative estimations of the effectiveness of quasiaprticle
traps.

The quasiparticles injected into superconductor by NIS junction
create a nonequilibrium distribution which is characterized by the
so called charge (or branch) imbalance. The number of
electron-like excitations in not equal to the number of hole-like
ones any more. This effect is known since 1972, when Clarke has
carried out first experiments on charge imbalance \cite{Clarke1}
with NIS junctions. Under the existence of charge imbalance the
simple formula (\ref{P0}) is not applicable and should be
modified. Below we will modify it applying a method proposed by
Tinkham \cite{CT,Tinkham}. Then we consider a model NISIN system
with two junctions in series (Fig.1). We assume that the
resistance of the first junction ($R_1$) is high, while the
resistance of the second one ($R_2$) is low. In this case the
first junction cools the left normal electrode, while the second
junction partially removes excited quasiparticles from the
superconductor. The right normal metal lead models the
quasiparticle trap \cite{Pekola2}. For the sake of simplicity we
assume that the junction number 2 is a tunnel junction  and
neglect proximity effect.

\section{Theory}

In this section we closely follow the method
proposed by Tinkham \cite{Tinkham}.

Let us first consider a single NIS junction. For the sake
of definiteness we choose the junction number 1.
The Hamiltonian of the system can be written as follows:
\begin{eqnarray}
\widehat H=\widehat H_N +\widehat H_S +\widehat H_T,
\end{eqnarray}
where $\widehat H_N$ is the Hamiltonian of normal metal,
$\widehat H_S$ is the Hamiltonian of superconductor and
$\widehat H_T$ is the tunnel Hamiltonian. They are given by the
following expressions:
\begin{eqnarray}
\widehat H_N&=&\sum_{k,\alpha}(\epsilon_k-eV_1)
c^+_{k,\alpha}c_{k,\alpha},
\nonumber\\
\widehat H_S&=&\sum_{n,\alpha}\left\{\xi_n-E_n+E_n\gamma^+_{
n,\alpha}\gamma_{n,\alpha}\right\},
\nonumber\\
\widehat H_T&=&\sum_{k,n,\alpha}\left\{t_{k n}c^+_{
k,\alpha}a_{n,\alpha} +t^*_{k
n}a^+_{n,\alpha}c_{k,\alpha}
\right\}
=\sum_{k,n,\alpha}\big\{t_{k n}c^+_{
k,\alpha}[u_n\gamma_{n,\alpha}-\alpha v^*_n\bar\gamma^+_{n,-\alpha}]
\nonumber\\
&&
+\;t^*_{k n}[u_n^*\gamma^+_{
n,\alpha}-\alpha v_n\bar\gamma_{n,-\alpha}]c_{k,\alpha}\big\}.
\end{eqnarray}
Here $V_1$ is the potential of the first normal metal lead
relative to the superconductor, $\epsilon_k$ is the energy of
electrons in the normal metal referred to the appropriate chemical
potential, $E_n=\sqrt{\xi^2_n+\Delta^2}$ is the energy of
quasiparticles in the superconductor and $\xi_n$ are the energies
of electrons in the superconductor ( $\xi_n$ has the same meaning
as $\epsilon_k$ in the normal electrode). The index $k$ enumerates
the states in the normal metal, $n$ does the same in the
superconductor, while $\alpha=\pm 1$ is the spin index. We have
also defined the creation $c^+_{k,\alpha}$ and annihilation
$c_{k,\alpha}$ electron operators in the normal metal, analogous
operators $a^+_{n,\alpha}$ and $a_{n,\alpha}$ in the
superconductor, and quasiparticle operators $\gamma^+_{n,\alpha}$
and $\gamma_{n,\alpha}.$ The latter operators are related to the
electron operators by means of the standard transformation
$a_{n,\alpha}=u_n\gamma_{n,\alpha}-\alpha
v_n^*\bar\gamma^+_{n,-\alpha},$ where the operator
$\bar\gamma^+_{n,-\alpha}$ creates a quasiparticle in the state
which is time reversed with respect to the $n$-th state and also
carries an opposite spin.  Finally, the BCS coherence factors are
given by the following standard expressions:
$|u_n|^2,\,|v_n|^2=\frac{1}{2}\left(1\pm\frac{\xi_n}{E_n}\right).$

We consider the tunneling Hamiltonian as a perturbation. The
current operator in interaction representation can be defined as
the rate of tunneling out of the normal metal multiplied by the
electron charge $-e$:
\begin{equation}
\widehat I=(-e)\bigg(-\frac{d }{dt}\sum_{k,\alpha}
c^+_{k,\alpha}c_{k,\alpha} \bigg)=ie\bigg[\widehat H_T,
\sum_{k,\alpha}
c^+_{k,\alpha}c_{k,\alpha} \bigg].
\end{equation}
The average value of the current can be evaluated by means of Fermi golden rule:
\begin{eqnarray}
I&=&\langle\widehat I(t)\rangle=i\int\limits_{-\infty}^t dt'\,
\langle [\widehat H_T(t'),\widehat I(t) ]\rangle
\nonumber\\
&=&4e\;{\rm Re}\sum_{k,n}|t_{kn}|^2\int\limits_{-\infty}^t dt'
\biggl[ |u_n|^2{\rm e}^{i(E_n-\epsilon_k+eV_1)(t-t')}(f_{S,n}
-f_{N,k})
\nonumber\\
&&
+ |v_n|^2{\rm e}^{-i(E_n+\epsilon_k-eV_1)(t-t')}(1-f_{N,k}-f_{S,n})
 \biggr].
\nonumber
\end{eqnarray}
Evaluating time integrals we obtain
\begin{eqnarray}
I&=&4\pi e\,\sum_{k,n}|t_{kn}|^2\biggl[\delta(E_n-\epsilon_k+eV)|u_n|^2
(f_{S,n}-f_{N,k})
\nonumber\\
&&
+\delta(E_n+\epsilon_k-eV)|v_n^2|(1-f_{N,k}- f_{S,n}) \biggr].
\end{eqnarray}
Now, as usual, we replace $|t_{kn}|^2$ by constant value $|t|^2,$ introduce densities of
states in both leads, assume the distribution functions depend
only on energy, and find
\begin{eqnarray}
I_{1}&=&\frac{1}{eR}\int
d\xi\left[\frac{1}{2}\left(1+\frac{\xi}{E}\right)(f_S(\xi)-f_N(E+eV_1))
\right.
\nonumber\\
&&\left.
+
\frac{1}{2}\left(1-\frac{\xi}{E}\right)(1-f_S(\xi)-f_N(-E+eV_1))
\right].
\label{INIS}
\end{eqnarray}
where $1/R=4\pi e^2|t|^2N_0^NN_0^S{\cal V}^N{\cal V}^S,$ and
${\cal V}^{N,S}$ are the volumes of the normal and superconducting
leads respectively. This result has been first derived by Tinkham
\cite{Tinkham}. Only if the distribution function in the
superconductor satisfies the particle-hole symmetry requirement
$f_S(-\xi)=f_S(\xi)$,  and  the distribution function in the
normal metal satisfies $f_N(-E)=1-f_N(E)$  Eq. (\ref{INIS})
reduces to the standard one:
\begin{equation}
I_{1}^{\rm sym}=\frac{1}{eR}\int dE\, \frac{\theta(E^2-\Delta^2)|E|}{\sqrt{E^2-\Delta^2}}
[f^S(E)-f^N(E+eV_1)].
\label{standard}
\end{equation}
Note that there exist a difference in definition between the distribution functions
$f_S(\xi)$ in (\ref{INIS}) and $f_S(E)$ in (\ref{standard}). In the first case the distribution
function depends on the energy of bare electrons in superconductor $\xi,$ while
in the second case --- on $E=\sqrt{\xi^2+\Delta^2}.$ Moreover, negative values of $E$
are allowed in (\ref{standard}), the distribution function for such values is defined
by means of the relation $f_S(-E)=1-f_S(E).$

The cooling power of the normal metal is obtained analogously.
First we note that $P_1=-\frac{d}{dt}\sum_{k,\alpha}\epsilon_k c^+_{k,\alpha}c_{k,\alpha},$
and then repeat all the steps of the previous procedure.
Thus we find
\begin{eqnarray}
P_{1}&=&\frac{1}{e^2R}\int
d\xi\left[\frac{E+eV_1}{2}\left(1+\frac{\xi}{E}\right)(f_N(E+eV_1)-f_S(\xi))
\right.
\nonumber\\
&&\left.
+
\frac{E-eV_1}{2}\left(1-\frac{\xi}{E}\right)(1-f_S(\xi)-f_N(-E+eV_1))
\right].
\label{PNIS}
\end{eqnarray}
Again we note that only if  $f_S(-\xi)=f_S(\xi)$  and  $f_N(-E)=1-f_N(E)$
the latter result reduces to a simple formula (\ref{P0}).

The same approach can be applied in order to find the rates of population
and depopulation of quasiparticle states in superconductor.
Here we already consider the system of two junctions.
We find
\begin{eqnarray}
 \frac{d f_{S,n}}{dt}&=&\frac{d}{dt}\langle\gamma^+_{n,\alpha}\gamma_{n,\alpha}\rangle
=-\frac{f_{S,n}-f_{0}(E_n)}{\tau_0}
\nonumber\\
&&
+\frac{|u_n|^2f_{N,1}(E_n+eV_1)
+|v_n|^2(1-f_{N,1}(-E_n+eV_1))-f_{S,n}}{2e^2R_1N_0^S{\cal V}^S}
\nonumber\\
&&
+
\frac{|u_n|^2f_{N,2}(E_n-eV_2)
+|v_n|^2(1-f_{N,2}(-E_n-eV_2))-f_{S,n}}{2e^2R_2N_0^S{\cal V}^S},
\label{kin}
\end{eqnarray}
where  $f_{N,1}$ and $f_{N,2}$ are the distribution functions is
the two normal metal electrodes, $f_0(E)=1/(\exp(E/T)+1),$ and
$\tau_0$ is the inelastic quasiparticle relaxation time in the
superconductor. We use a very simple relaxation term in Eq.
(\ref{kin}) because we actually do not consider the effect of
quasiparticle interactions here. The relaxation term is introduced
in the Eq. (\ref{kin}) only in order to establish the limitations
of a non-interacting model. The second term in the right hand side
of Eq. (\ref{kin}) describes the injection of quasiparticles
through the first junction, while the last term -- the injection
through the second junction. The equations similar to Eq.
(\ref{kin}) are derived and discussed in Ref. \cite{Pethick,Clarke2,Heslinga}.

In a stationary case, $\dot f_n^S\equiv 0,$ the distribution function
of quasiparticles in the superconductor can be found explicitly:
\begin{eqnarray}
f_S(\xi)&=&\frac{1}{{\frac{N_0^S{\cal V}^S}{\tau_0}+
\frac{1}{2e^2R_1}+\frac{1}{2e^2R_2}}}
\left\{\frac{N_0^S{\cal V}^S}{\tau_0}f_0(E)
\right.
\nonumber\\
&&
\left.
+\,\frac{1}{2e^2R_1}
[|u|^2f_{N,1}(E+eV_1)+|v|^2f_{N,1}(E-eV_1)]
\right.
\nonumber\\
&&\left.
+\,\frac{1}{2e^2R_2}
[|u|^2f_{N,2}(E-eV_2)+|v|^2f_{N,2}(E+eV_2)]\right\}.
\label{fS}
\end{eqnarray}
Here we have assumed that the distribution functions in the normal
metal electrodes satisfy $f^N_{1,2}(-E)=1-f^N_{12}(E).$
The result (\ref{fS}) is similar to that obtained in Ref. \cite{Heslinga},
but it includes the effects of charge imbalance which  become important
if the resistances $R_1$ and $R_2$ are not equal.

The shape of the quasiparticle distribution function (\ref{fS})
depends on two parameters: the junction asymmetry parameter
$\alpha=R_2/(R_1+R_2)$ and the ratio of inelastic relaxation rate
to the injection rate $ \beta={2e^2R_1 N_0^S{\cal V}^S}/{\tau_0}.
$ One can neglect  inelastic relaxation as long as $\beta\ll 1.$
Here we have also assumed that quasiparticle distribution function
does not depend on space coordinates. This assumption is valid if
the appropriate size of the superconducting electrode does not
exceed the quasiparticle relaxation length
$L_{in}=\sqrt{D\tau_0},$ where $D$ is the diffusion constant.

Finally, the superconducting gap is determined by BCS  gap equation:
\begin{equation}
1=\frac{\lambda}{2}\int d\xi\,\frac{1-2f_S(\xi)}{\sqrt{\xi^2+\Delta^2}},
\label{BCS}
\end{equation}
where $\lambda$ is the electron-phonon coupling constant.

\section{Results and discussion}

\begin{figure}
\begin{center}
\includegraphics[width=9cm]{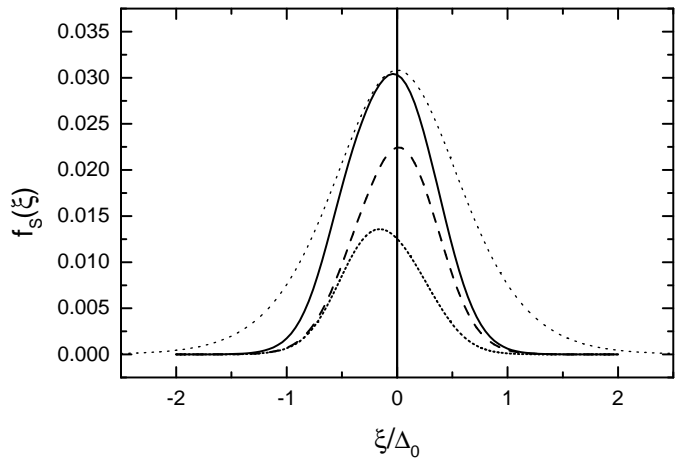}
\end{center}
\caption{Quasiparticle distribution function in the superconductor
for different values of the parameters. (1)~Dotted line --
equilibrium distribution function $f_0(\xi)=1/[1+\exp(E/T_{eff})]$
with effective temperature $T_{eff}/\Delta_0=0.3.$ (2)~Solid line
-- bias voltage is applied between normal metals (Fig. 1a),
$eV/\Delta_0=1.72,$ $T/\Delta_0=0.1,$ $\alpha =0.05,$ no inelastic
relaxation ($\beta=0$). (3)~Dashed line -- all parameters have the
same values as in the previous case, but the inelastic relaxation
time is finite, $\beta=10.$ (4)~Short dot -- bias voltage is
applied between normal metal and superconductor (Fig. 1b),
$eV/\Delta_0=1,$ $T/\Delta_0=0.1,$ $\alpha=0.05,$ $\beta=0.$}
\end{figure}

We have solved the set of equations (\ref{INIS},\ref{PNIS},\ref{fS}) numerically.
It has been done as follows.
First, we have chosen the equilibrium distribution functions with the same
temperature $T$ in both normal metal leads.
If the bias voltage is applied between the normal metals,
(Fig. 1a), then
 the voltages across
individual junctions are fixed by the current conservation requirement
$I_1(V_1)=I_2(V_2)$ and $V_1+V_2=V.$ These equations
are solved numerically.  If only the junction number 1 is biased (Fig. 1b),
then we just set $V_1=V,$ $V_2=0.$  As long as $V_1$ and $V_2$ are
known, we find the distribution function
(\ref{fS}) and the cooling power (\ref{PNIS}).
Below we present some results.
All energy scales are normalized by the superconducting
gap at zero temperature, $\Delta_0\equiv \Delta(T=0).$

Fig. 2 shows the distribution function for different sets of the
parameters. In all cases this function has an asymmetry with
respect to the chemical potential of Cooper pairs, this a
manifestation of charge imbalance. This effect turns out to be
relatively weak if the bias voltage is applied between the normal
metals. Once again we note that the distribution function does not
look like a Fermi function because of its definition (see the
discussion after Eq. (\ref{standard})).

\begin{figure}
\begin{center}
\includegraphics[width=11.5cm]{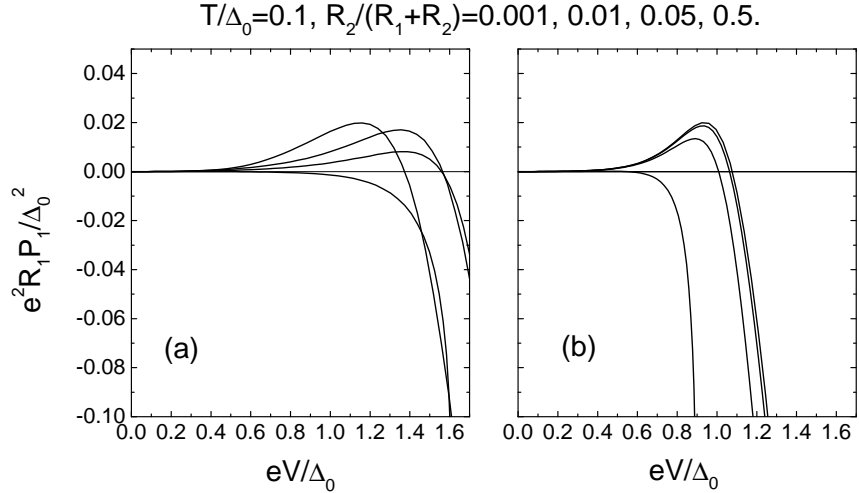}
\end{center}
\caption{Cooling power for different junctions asymmetry and in
the absence of the inelastic relaxation. (a)~The voltage is
applied between two normal metals (Fig. 1a). (b)~The voltage is
applied between the normal metal and the superconductor (Fig.
1b).}
\end{figure}
The cooling power of the first junction  (\ref{PNIS}) as function
of voltage is shown in Fig. 3. We have neglected inelastic
processes in superconductor assuming $\beta\ll 1.$ We observe that
even a relatively small resistance of the second junction may
significantly reduce the cooling power. It is quite natural
because the cooling of the superconductor becomes less effective.
In the opposite case $\beta\gg 1$ the cooling power is given by
the standard formula (\ref{P0}). However, in this case the power
dissipated in the superconductor goes first to the phonon
subsystem, and then may return back to the cooled normal metal
\cite{Jochum}. This effect may strongly reduce the overall cooling
power of the device. In the absence of inelastic relaxation such
processes are less probable, because power dissipation takes place
in the normal metal with high heat conductivity and the heat is
effectively removed from the device. That's why below we
consider the case $\beta\ll 1.$

\begin{figure}
\begin{center}
\includegraphics[width=9.5cm]{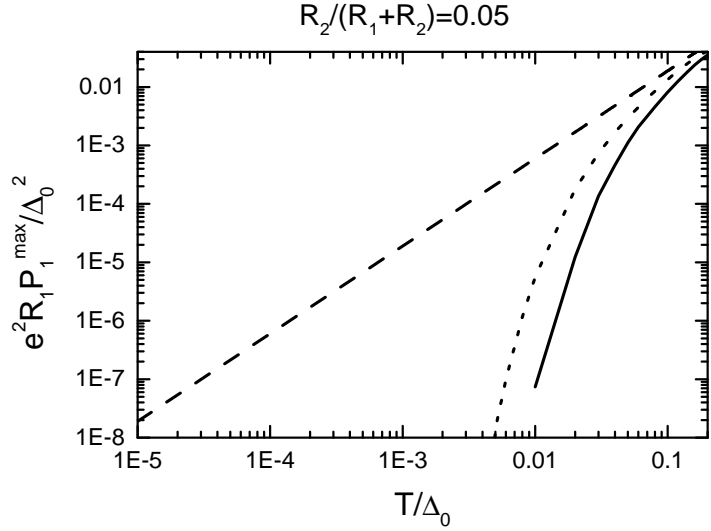}
\end{center}
\caption{Dependence of maximum cooling power on temperature.
Inelastic relaxation in superconductor is neglected. Solid line --
maximum cooling power at $\alpha=0.05$ and bias voltage applied
between the normal metal leads (Fig. 1a). Dotted line -- maximum
cooling power at $\alpha=0.05$ and bias voltage applied to the
first junction (Fig 1b). Dashed line -- equilibrium result based
on Eq. (\protect\ref{P0}),
$P_1^{\max}=0.6({\Delta_0^2}/{e^2R_1})({T}/{\Delta_0})^{3/2}$
\protect\cite{Pekola1}.}
\end{figure}

\begin{figure}
\begin{center}
\includegraphics[width=9.5cm]{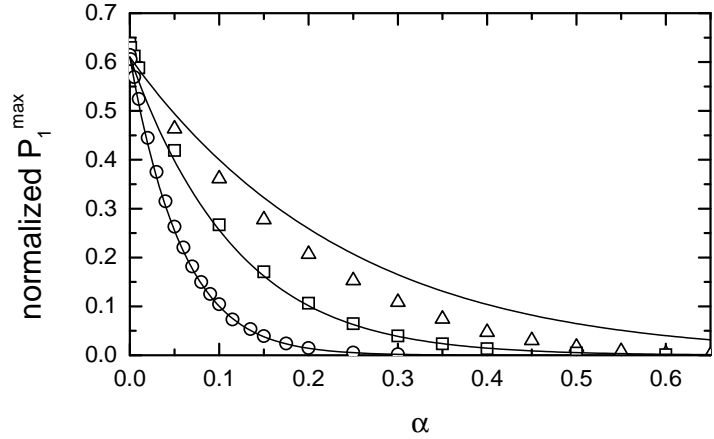}
\end{center}
\caption{Dependence of maximum cooling power on asymmetry
$\alpha.$ $\beta=0,$ voltage is applied to the first junction
(Fig. 1b). The cooling power is normalized by
$({\Delta_0^2}/{e^2R_1})({T}/{\Delta_0})^{3/2}.$ Triangles,
squares and circles --- numerical results for $T/\Delta_0=0.2,\;
0.1,\; 0.05$ respectively. Solid lines --- simple approximation
(\protect\ref{Pmax}). }
\end{figure}
The dependence of the maximum cooling power on temperature is
illustrated in Fig. 4.  The cooling power is drastically reduced
at low temperatures $T/\Delta_0<\alpha.$  Fig. 5 shows the
dependence of the maximum cooling power on the asymmetry $\alpha.$
Here we again observe the reduction of the cooling power for
$\alpha>T/\Delta_0,$ in agreement with the Fig. 4. If the bias
voltage is applied only to the first junction and the temperature
is much lower than $\Delta_0$, one can get an approximate
analytical expression for the maximum cooling power:
\begin{equation}
P_{\rm 1,max}=\frac{\Delta_0^{1/2}T^{3/2}}{e^2R_1}
\left[
0.76{\rm e}^{-\left(\frac{\Delta_0}{T}-\frac{1}{8} \right)\alpha}-
0.22{\rm e}^{-2\left(\frac{\Delta_0}{T}-\frac{1}{8} \right)\alpha}
+0.07{\rm e}^{-3\left(\frac{\Delta_0}{T}-\frac{1}{8} \right)\alpha}
\right].
\label{Pmax}
\end{equation}
If the voltage is applied to the normal metal electrodes, the
maximum cooling power is even lower than (\ref{Pmax}).

\begin{figure}
\begin{center}
\includegraphics[width=9.5cm]{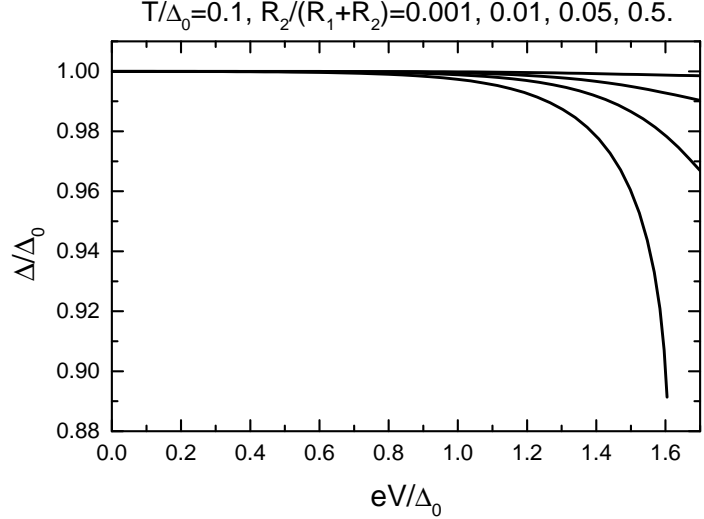}
\end{center}
\caption{The suppression of superconducting gap at different
degrees of asymmetry. The bias voltage is applied to the normal
metal electrodes (Fig. 1a).}
\end{figure}
Fig.~6 shows the suppression of the gap due to the heating of the
superconductor.

The model employed above requires two conditions to be satisfied:
the size of the superconductor should be less than $L_{in},$
and  $\beta$ should be small, $\beta\ll 1.$
Experimentally one finds $L_{in}$
to be of the order of at least several microns (see e.g. \cite{Pekola2,Ullom}).
Thus, the first requirement is not very restrictive, although it may depend
on sample geometry. The parameter $\beta$ is difficult to estimate because
the inelastic relaxation time is unknown. Most probably this time is of
the order of several microseconds (see e.g. \cite{Kaplan}). If so, then
the condition $\beta\ll 1$ can also be easily satisfied in experiment.

In conclusion, we have derived a generalized expression for the
cooling power of a NIS tunnel junction taking into account charge
imbalance effects. We have considered cooling properties of NISIN
double junction structure. The interactions in the superconductor
were neglected. Such an approximation is applicable provided the
superconducting electrode is of submicron size. It is shown that
the cooling power depends strongly on the ratio of the resistances
of the two junctions. We have shown that at low temperatures the
maximum cooling power of NISIN structure is proportional to
$\exp(-\Delta_0 R_2/T(R_1+R_2)).$

\end{document}